\documentclass[fleqn,usenatbib,letters]{mnras}

\usepackage[T1]{fontenc}
\usepackage{ae,aecompl}

\DeclareRobustCommand{\VAN}[3]{#2}
\let\VANthebibliography\thebibliography
\def\thebibliography{\DeclareRobustCommand{\VAN}[3]{##3}\VANthebibliography}

\usepackage{graphicx}	
\usepackage{amsmath}	
\usepackage{amssymb}	

\usepackage{amsfonts}
\usepackage{multicol}
\usepackage[normalem]{ulem}
\usepackage{bm}
\usepackage{color}
\usepackage{newtxtext,newtxmath}

\definecolor{grey}{rgb}{0.75,0.75,0.75}
\definecolor{Orange}{rgb}{1.0,0.5,0.15}
\definecolor{brown}{rgb}{0.7,0.25,0.0}
\definecolor{pink}{rgb}{1.0,0.5,0.5}
\definecolor{darkerred}{rgb}{0.8,0,0}
\definecolor{darkerblue}{rgb}{0,0,0.8}
\definecolor{Blue}{rgb}{0,0.08,0.65}
\definecolor{Red}{rgb}{0.65,0.08,0.05}
\definecolor{Green}{rgb}{0.15,0.45,0.25}

\newcommand{\rcm}{\bm{r}_{\mathrm{cm}}}
\newcommand{\cm}{\mathrm{cm}}

\newcommand{\dd}{\mathrm{d}}

\newcommand{\rr}{\bm{r}}

\newcommand{\Tr}{\mathrm{Tr}}

\newcommand{\tr}{\mathrm{tr}}
\newcommand{\mat}[1]{\bm{\mathsf{#1}}}

\title[Minimum energy principle]{Getting in shape with minimal energy. A variational principle for protohaloes}

\author[M. Musso and R. K. Sheth]{
Marcello Musso$^{1}$\thanks{E-mail: mmusso@usal.es}
and Ravi K. Sheth,$^{2,3}$\thanks{E-mail: shethrk@upenn.edu}
\\
$^{1}$Departamento de F\'{\i}sica Fundamental and IUFFyM,
Universidad de Salamanca, E-37008 Salamanca, Spain\\
$^{2}$Center for Particle Cosmology, University of Pennsylvania, 209 S. 33rd St., Philadelphia, PA 19104, USA\\
$^{3}$The Abdus Salam International Center for Theoretical Physics, Strada Costiera, 11, Trieste 34151, Italy
}

\date{Accepted XXX. Received YYY; in original form ZZZ}

\pubyear{2022}

\begin{document}
\label{firstpage}
\pagerange{\pageref{firstpage}--\pageref{lastpage}}
\maketitle

\begin{abstract}
In analytical models of structure formation, protohalos are routinely assumed to be peaks of the smoothed initial density field, with the smoothing filter being spherically symmetric.  This works reasonably well for identifying a protohalo's center of mass, but not its shape. To provide a more realistic description of protohalo boundaries, one must go beyond the spherical picture. We suggest that this can be done by looking for regions of fixed volume, but arbitrary shape, that minimize the enclosed energy. Such regions are surrounded by surfaces over which (a slightly modified version of) the gravitational potential is constant. We show that these equipotential surfaces provide an excellent description of protohalo shapes, orientations and associated torques.
\end{abstract}




\section{Introduction}

The key ingredient of most analytical models of structure formation is the initial (over-)density field $\delta$ smoothed with a suitable spherical filter.  The spherical collapse model \citep{gg72} establishes a mapping between the initial enclosed density and the collapse time, leading to a critical initial mean density $\delta_c$ that a region must have for it to collapse at the present time.  Motivated by this, the earliest models tried to characterize protohalo regions as spheres of a prescribed initial mean density \citep{ps74}. Subsequent work noticed that if a sphere of critical mean density is enclosed within a larger one of the same density, then the protohalo should be identified with the larger volume.  Hence, to avoid double counting, a protohalo should be assigned the mass contained in the \emph{largest} sphere of critical mean density centered on that location \cite[the so-called \emph{cloud-in-cloud} problem of][]{bcek91}. 

Other models, focusing more on the locations of protohalos rather than their formation times, emphasized that collapse does not take place around random locations \citep{smt01}.  For instance, spheres that are denser than their neighbors collapse faster, so it is reasonable to identify protohalo centers with the centers of spheres that are local maxima of the mean density \cite[the \emph{peaks theory} of][]{bbks86}. The two approaches have been combined together, leading to the description of protohalos as peaks of the smoothed density field that have critical height, and are not contained within other peaks of the same height but larger smoothing scale \cite[the \emph{excursion set peaks} approach: ][]{jrb89,aj90,ps12,psd13}. 

Considerable attention has also been devoted to the precise value of the critical value for collapse and the functional form of the filter. The former, leading to more realistic thresholds that are both stochastic (i.e. dependent on variables other than the mean overdensity) and scale dependent, and can include the anisotropy of the environment \citep{bm96,smt01,cl2001}. The latter, in order to avoid the mathematical drawback of divergences arising with the usual top-hat filter \citep[e.g. the \emph{CUSP} approach of][and references therein]{cusp2021}, but also to deal with more physically motivated quantities such as the protohalo's energy, rather than mass density \citep{epeaks}.

Although halos in simulations are not usually spherically symmetric \citep[][and references therein]{bonamigo2015}, and the protohalo patches from which they formed are not either \citep{dts13, porciani11, otBAO}, to the best of our knowledge, most analytical studies to date have used a spherical volume to model protohalos \cite[but see][for initially nonspherical regions that evolve to become spheres]{cza02, ls08, borzyShear}.
Because protohalos are not spherical, this can, at best, identify the center of mass.  Although suitable averages with this filter can offer clues to the actual protohalo shape \citep{bm96,monaco97} and its subsequent evolution \citep{ecShapes, cusp2021}, their accuracy is limited \citep{porciani11}, in part because they do not provide a physical handle to describe the protohalo boundary. 

In this Letter, we assume no \emph{a priori} shape, but suggest that both the location \emph{and} the shape of a protohalo are associated with a region of minimal initial energy, whose boundary is asurface on which (a slightly modified form of) the gravitational potential perturbation remains constant. This equipotential surface is the locus of points where the initial infall velocity (the component of the velocity directed towards the center of mass) is constant. Section~\ref{sec:concept} describes the minimum energy principle in detail. 
Section \ref{sec:sims} tests this \emph{ansatz} against protohaloes in an N-body simulation.
A final section summarizes.

\section{Conceptual foundation}\label{sec:concept}
This section makes two conceptual points: 
 (i) the shape that best describes a protohalo of a given mass is the one that, while preserving the volume, minimises the enclosed energy, and 
(ii) the boundary of this special region is an `equipotential' surface. 

\subsection{The miniumum energy principle}\label{sec:theory} 

In an earlier paper on this subject, \cite{epeaks} showed that the center of mass of a protohalo typically coincides with the center of a sphere which contains the same mass $M$ and maximises the \emph{potential energy overdensity}
\begin{align}
  \epsilon &\equiv
  \frac{5}{MR_I^2}\int_V\dd\rr\, \rho(\rr)\,(\rr-\rcm)\cdot[\nabla\phi(\rr) - \nabla\phi_\cm] 
\label{eq:epsilon}
\end{align}
where $\phi$ is the potential perturbation, normalized so that $\nabla^2\phi = \delta$, 
$\rcm$ and $-\nabla\phi_\cm$ are the center of mass position and acceleration of the volume $V$, and 
$R_I$ is its \emph{inertial radius}, defined as
\begin{equation}
  R_I^2\equiv \frac{5}{3M}\int_V\dd\rr\, \rho(\rr) \,|\rr-\rcm|^2\,.
\label{eq:RI}
\end{equation}
Although the definitions above are general, they simplify in the initial conditions, when $\rho(\rr)\simeq\bar\rho$, $M\simeq\bar\rho V$ and $\nabla\phi\simeq - \bm{v}/fDH$, where $\bm{v}$ is the initial velocity and $fHD\equiv (d\ln D/d\ln a)(d\ln a/dt)D = dD/dt$, where $D$ is the linear theory growth rate. Furthermore, either $\nabla\phi_\cm$ or $\rcm$ in equation~\eqref{eq:epsilon} may be omitted, since the mass averages of both $\rr-\rcm$ and of $\nabla\phi-\nabla\phi_\cm$ vanish by definition.

For a homogeneous sphere, $R_I$ coincides with the sphere's geometrical radius $R$ (defined by $V=4\pi R^3/3$), and $\epsilon$ with its mean matter overdensity $\delta_R$. In general, however, they are different. For spheres centered on protohalos, which tend to be more centrally overdense, usually $R_I<R$ (with the difference starting at first order in perturbations) and $\epsilon>\delta_R$. 
At early times, the potential energy overdensity is related to the total energy $E$ (kinetic plus potential) within $V$ as
\begin{equation}
  E = - \frac{4\pi G}{3}\bar\rho\, M R_I^2\,\epsilon
    = -\frac{GM^2}{R}\, \frac{R_I^2}{R^2}\,\epsilon\,,
\label{eq:E}
\end{equation} 
This is $(5\epsilon/3)(R_I/R)^2$ times the potential energy of a homogeneous sphere containing the same mass as $V$, that is $-(3/5)\,(GM^2/R)$.  

The dynamical role of $\epsilon$ is twofold. First, it represents for $R_I$ what $\delta_R$ represents for $R$, that is, it determines the time at which $R_I$ turns around, according to the spherical collapse model: $R_I$ of a patch with larger $\epsilon$ shrinks faster, and that patch collapses sooner. Thus, the inertial radius of a volume $V$ that is a local maximum in $\epsilon$ has a smaller collapse time than all neighbouring volumes of the same mass. Secondly, for a sphere, the spatial gradient $\nabla\epsilon$ is (proportional to) the sphere's dipole moment. If this gradient vanishes, then so does the dipole term of the gravitational potential at the surface. Hence, if higher order multipoles are neglected, the initial velocities of the particles at the surface all approximately converge to the center. For these reasons, local maxima of the smoothed energy overdensity field are excellent candidates for the centers of mass of initial patches that will evolve dynamically into high density regions: i.e., protohaloes.

So far, like most of the literature to date, we have implicitly assumed $V$ to be a sphere. 
However, different shapes of the same $V$ will generically have different values of $\epsilon$. In particular, if one deforms a sphere's boundary so as to include larger values of $(\rr-\rcm)\cdot\nabla\phi$ and exclude smaller ones, while keeping the total mass fixed, then the new value of $\epsilon$ will be larger.  Thus, if one allows for such non-spherical deformations around an initially spherical peak, one can always find a value of $\epsilon$ that exceeds that of the spherical peak. However, any such deformation also changes $R_I^2$; sufficiently large deformations may increase the denominator of $\epsilon$ more than the numerator. 
 There is therefore a well defined, non-spherical surface that maximises the value of $\epsilon$. As the inertial radius $R_I$ of the region within this surface shrinks even faster than the sphere's, it is tempting to take this (non-spherical) region as the best candidate for the protohalo patch.


\begin{figure}
  \includegraphics[width=.9\columnwidth]{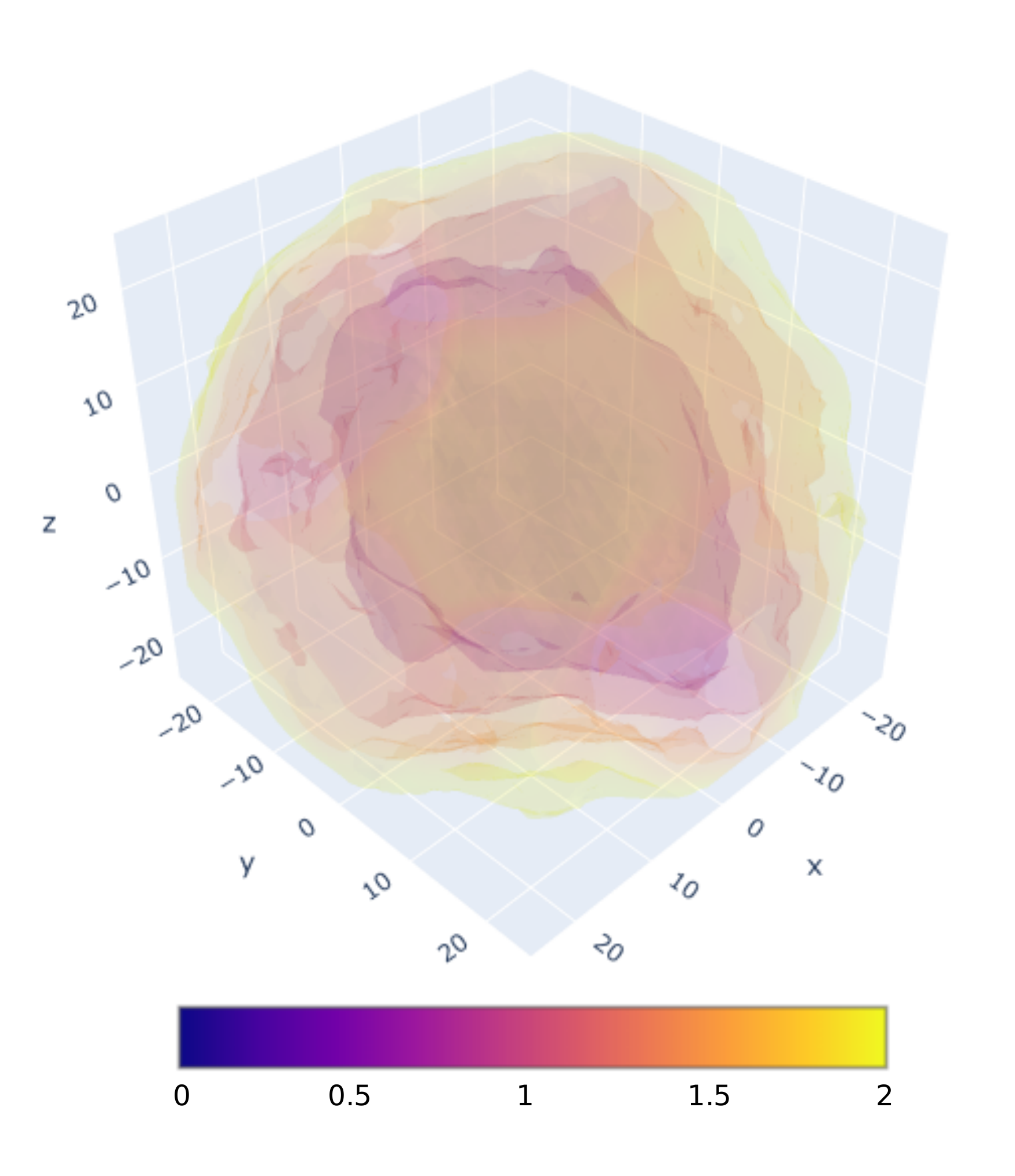}
\caption{Isosurfaces of the infall potential $\mathcal{V}/R_I^2$, defined in equation \eqref{eq:f}, centered on a protohalo patch. The values of $\epsilon$, $\rcm$ and $\nabla\phi_\cm$ are the same (the ones of the protohalo) for all surfaces. 
}
\label{fig:isosurfaces}
\end{figure}

\subsection{Equipotential surfaces}
\label{sec:equipot}

We now want to characterize the surface that maximises $\epsilon$. Since we just argued that there is always such a surface surrounding a spherical peak, it should be sufficient to find that surface for which the infinitesimal variation $\delta\epsilon$ vanishes for \emph{any} further infinitesimal deformation that preserves the mass.


The variation of the mass $M$ under infinitesimal deformations of the volume $V$ is given by the surface integral
\begin{equation}
  \delta M = \int_S \dd\bm{S}\cdot\delta\bm{\lambda} \, \rho(\rr)
\label{eq:dM}
\end{equation}
where $S$ is the boundary surface of $V$, $\dd\bm{S}=\dd S\,\bm{\hat n}$ is the normal to the surface element, and $\delta\bm{\lambda}$ is the infinitesimal deformation vector mapping each point on $S$ to its image. One must have $\delta M = 0$ for the deformation to preserve the mass. 

\begin{figure}
  \includegraphics[width=.9\columnwidth]{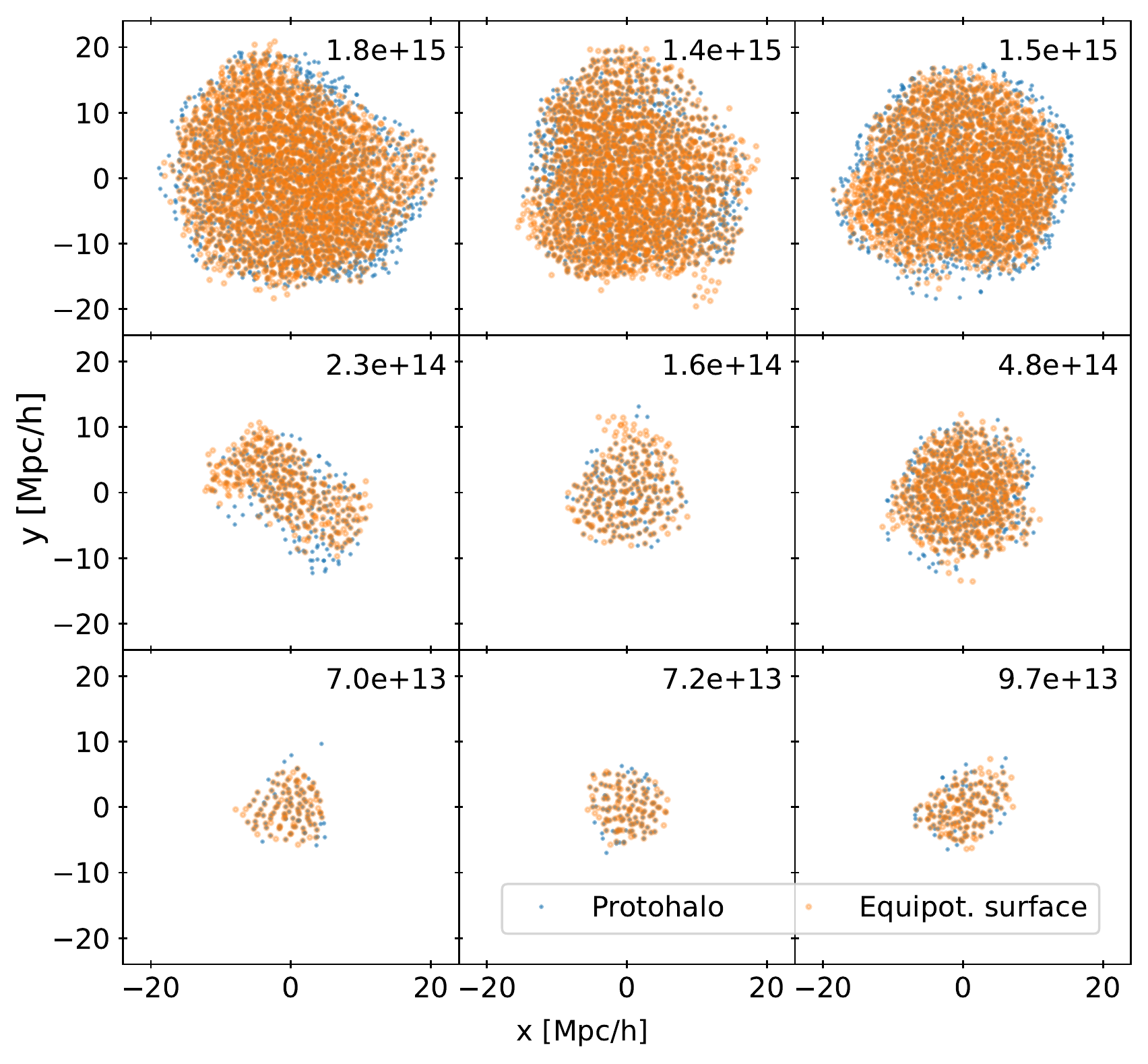}
\caption{\label{fig:protohalos} Comparison of protohalo patches (blue) with their equipotential regions  (orange; i.e., the region enclosed by isocontours of $\mathcal{V}$).  In all cases, particles are shown in the $x$-$y$ plane, projected in the $z$ direction.  Top right corner of each panel gives the protohalo mass (in $h^{-1}M_\odot$).}
\end{figure}

To compute the total variation of $\epsilon$, one must not only vary the integral over $V$ appearing in equation \eqref{eq:epsilon} explicitly, but also the one in $R_I^2$ (varying those in $\rcm$ and $\nabla\phi_\cm$ gives a null net contribution). Using Leibnitz's rule, one gets
\begin{equation}
  \delta\epsilon = -\frac{5}{MR_I^2}
  \int_S \dd\bm{S}\cdot\delta\bm{\lambda} \,\rho(\rr) \mathcal{V}(\rr)\,,
\label{eq:deps}
\end{equation}
where
\begin{equation}
  \mathcal{V}(\rr) \equiv (\rr-\rcm)\cdot\bigg[-(\nabla\phi-\nabla\phi_\cm)
  +\frac{\epsilon}{3} (\rr-\rcm)\bigg]
\label{eq:f}
\end{equation}
is the radial component of the acceleration relative to the center of mass, minus its mass weighed average over $V$. It has the dimensions of, and for the monopole term it actually equals, the potential relative to the center of mass minus its average\footnote{Near the protohalo boundary $\nabla\phi-\nabla\phi_\cm \sim\delta_R(\rr-\rcm)/3$ and $|\rr-\rcm|\sim R\sim R_I$, so we expect $\mathcal{V}/R_I^2\sim\epsilon-\delta_R$ to be of order unity \cite[e.g][]{epeaks}.}. With a slightly stretched terminology, we dub $\mathcal{V}(\rr)$ the \emph{infall potential}.

To make $\epsilon$ stationary as per our original aim, we need to find the surface $S$ such that $\delta\epsilon =0$ under \emph{any} deformation that preserves the mass. By comparing equations \eqref{eq:dM} and \eqref{eq:deps}, it follows immediately that this must be the surface over which $\mathcal{V}(\rr)$ is constant, since in this case one has $\delta\epsilon\propto\delta M = 0$. That is, $S$ is an iso-surface of $\mathcal{V}$; and, to the extent that $\mathcal{V}$ can be related to the gravitational potential, an equipotential surface.

There will be of course many of these surfaces, usually nested in one another, with different values of $\mathcal{V}$. Only one of them will contain exactly the mass $M$. Increasing (or decreasing) $M$ will select a different surface, usually enclosing (or enclosed by) the previous one. Each surface will also have its own value of $\epsilon$, which is maximal at fixed $M$ but not necessarily as $M$ varies. Just like in plain excursion sets, this value can be associated to the time at which the mass $M$ is assembled. The set of these nested iso-surfaces will thus provide a description of the halo's mass accretion history.

If $(\nabla\phi-\nabla\phi_\cm)$ had only the monopole term, then $\mathcal{V}(\rr)$ would be spherically symmetric, and all its iso-surfaces would be spheres centered on $\rcm$. In general, the presence of higher multipoles deforms them. Since $-(\rr-\rcm)\cdot(\nabla\phi-\nabla\phi_\cm)$ is proportional to the infall velocity, and normally becomes more negative as the distance grows, the boundary is pushed farther where the infall velocity is larger than its spherically symmetric value. Particles that are infalling faster will make it into the halo from further away. Thus, the minimum energy principle describes the asphericity of protohalos as a response to the anisotropy of the gravitational infall.

\section{Measurements in simulations}
\label{sec:sims}

We now test our minimum energy \emph{ansatz} using protohaloes from the Flora simulation, the largest box in the SBARBINE suite \citep{despali16}. The simulation evolved $1024^3$ dark matter particles each of mass $6.35\times10^{11} h^{-1}M_\odot$ in a periodic cube of side $L_{\mathrm{box}}=2h^{-1}$Gpc with a Planck13 background cosmology: $\Omega_m = 0.307$, $\Omega_\Lambda = 0.693$, $\sigma_8 = 0.829$ and $h = 0.677$.

Our halo set contains 5378 haloes identified at $z=0$ using a Spherical Overdensity (SO) halo finder with threshold of $319\times$ the background density. The set includes all the 1378 haloes more massive than $10^{15}h^{-1}M_\odot$, 2000 randomly chosen haloes with masses between $10^{14}$ and $10^{15}h^{-1}M_\odot$, and 2000 randomly chosen haloes with masses between $4\times 10^{13}$ and $10^{14}h^{-1}M_\odot$.
We refer to the patch defined by each halo's particles in the initial conditions as the `protohalo'.  
For a subset of haloes (the 1387 most massive halos, and 1372 in the intermediate mass bin) we also have counterparts identified with an Ellipsoidal Overdensity (EO) halo finder (typically slightly less round, and about 10 percent more massive). We used this control set to check that our results do not depend strongly on the halo finder. To emphasize the connection to energy peaks, we present results using  
 $\sigma_{02}^2(R)\equiv \int dk\,k^2P_{\rm Lin}(k)\,W_2^2(kR)$, 
 where $R \equiv (3M/4\pi\bar\rho)^{1/3}$ and $W_2(x)\equiv 15\,j_2(x)/x^2$; 
 increasing $M$ decreases $\sigma_{02}$.

\begin{figure}
  \includegraphics[width=0.95\columnwidth]{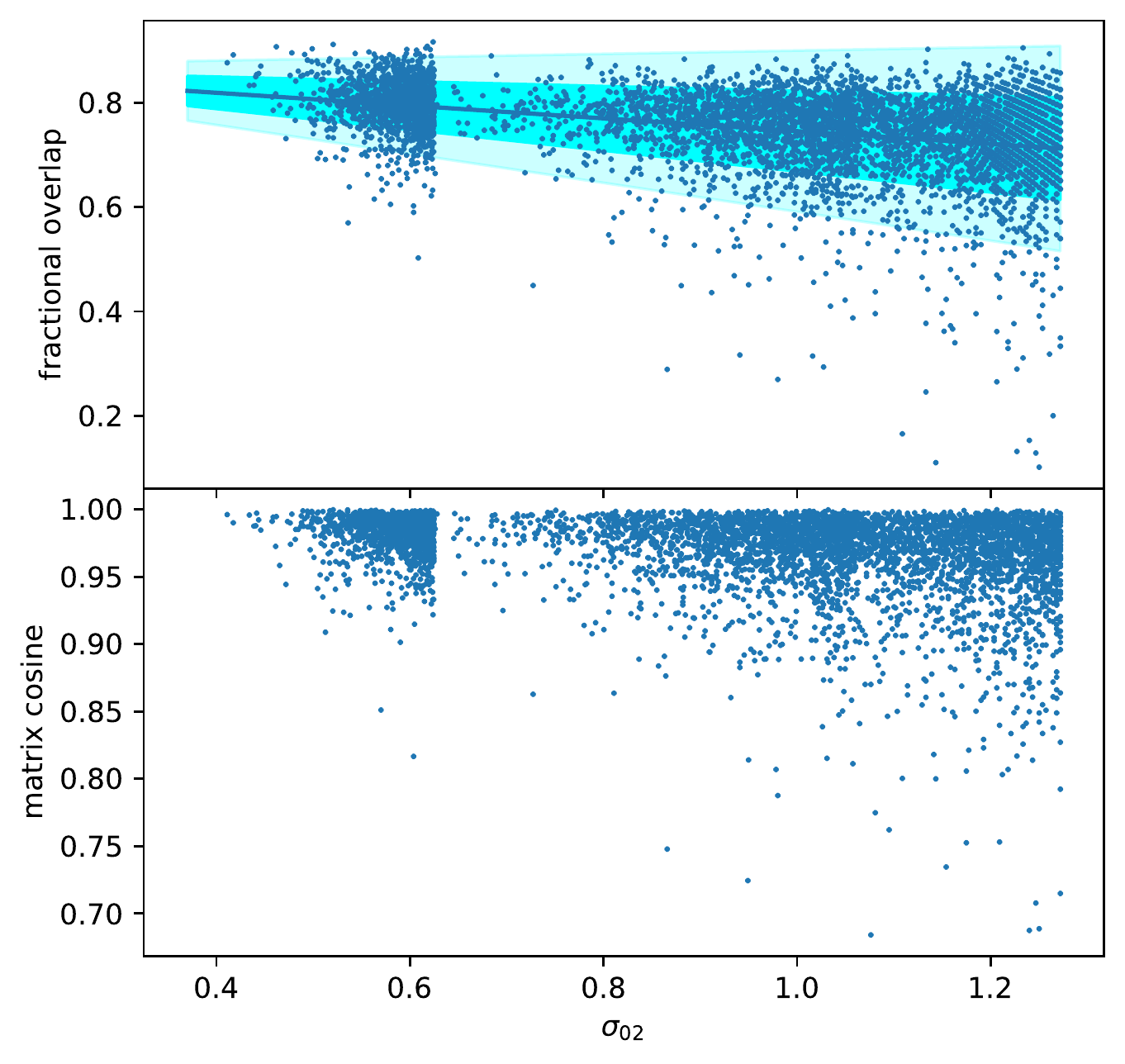}
  \caption{\label{fig:cosines} \emph{Top:} Fraction of actual protohalo particles included in the equipotential region.  
  \emph{Bottom:} Alignment between the inertia tensor of the protohalo and of the equipotential region (equation~\eqref{eq:cosine}). Both are shown as a function of $\sigma_{02}$ (largest masses on the left)}
\end{figure}

We measure each protohalo's center of mass position and velocity, $\rcm$ and $\bm{v}_\cm$, by averaging over all its particles. We then estimate its potential energy overdensity \emph{tensor} as
\begin{equation}
  \hat\epsilon_{ij} \equiv -3\, \frac{\sum_n [(\bm{r}-\rcm)_i(\bm{v} -
  \bm{v}_\cm)_j/fDH]_n}{\sum_n [(\bm{r}-\rcm)\cdot(\bm{r} - \rcm)]_n} \,,
\label{eq:eij}
\end{equation}
where $D$ is the $\Lambda$CDM density perturbation growth factor, $f=\dd\ln D/\dd\ln a$ so that $\dot D = fDH$, and $n$ runs over all $N$ particles in the protohalo. Next, with these values of $\rcm$, $\bm{v}_\cm$ and $\hat\epsilon =\tr(\mat{\hat\epsilon})$, we construct
\begin{equation}
  \hat{\mathcal{V}}(\rr) \equiv
  (\rr-\rcm)\cdot\bigg[\frac{(\bm{v} - \bm{v}_\cm)}{fDH} + \frac{\hat\epsilon}{3} (\rr-\rcm)\bigg]
\label{eq:estf}
\end{equation}
for each particle in a cube of side $3R$ centered on $\rcm$, and select the $N$ particles having the lowest values.
This singles out a region that has the same mass as the protohalo, and is bound by an iso-surface of $\mathcal{V}$, which we call the protohalo's \emph{equipotential region}. 

Fig. \ref{fig:isosurfaces} shows, for illustration, a few iso-surfaces of $\mathcal{V}$ (scaled by $R_I^2$ to make it dimensionless) for a protohalo in the largest mass bin, all relative to the \emph{same} values of $\rcm$, $\bm{v}_\cm$ and $\hat\epsilon$ (the ones of the protohalo). We refer to the one that encloses the \emph{same} number of particles (hence, the same mass) as the protohalo as the `boundary' of its equipotential region. As argued in the previous section, this region has $\mathcal{V}/R_I^2\sim 1$ and the maximal value of $\epsilon$ for the given mass, and is the most natural prediction for the protohalo boundary.
Deep inside the region, the equipotential surfaces break up into disconnected regions (not shown), which we believe encodes the assembly history. To make this connection stronger, however, one should not use the same  $\rcm$, $\bm{v}_\cm$ and $\hat\epsilon$ for the whole set, but recompute them for each surface at every step.

Figure \ref{fig:protohalos} compares the protohalo and equipotential regions for a few randomly selected haloes in the three mass bins.  Smaller blue dots represent protohalo particles, and larger orange circles the ones of the equipotential region, plotted in the $x$-$y$ plane and projected along the $z$ direction.  By visual inspection one can appreciate the excellent agreement between the two. It is remarkable how our approach has captured strongly non-spherical features like the lobes and protruding arms of most protohaloes.

\begin{figure}
  \includegraphics[width=0.95\columnwidth]{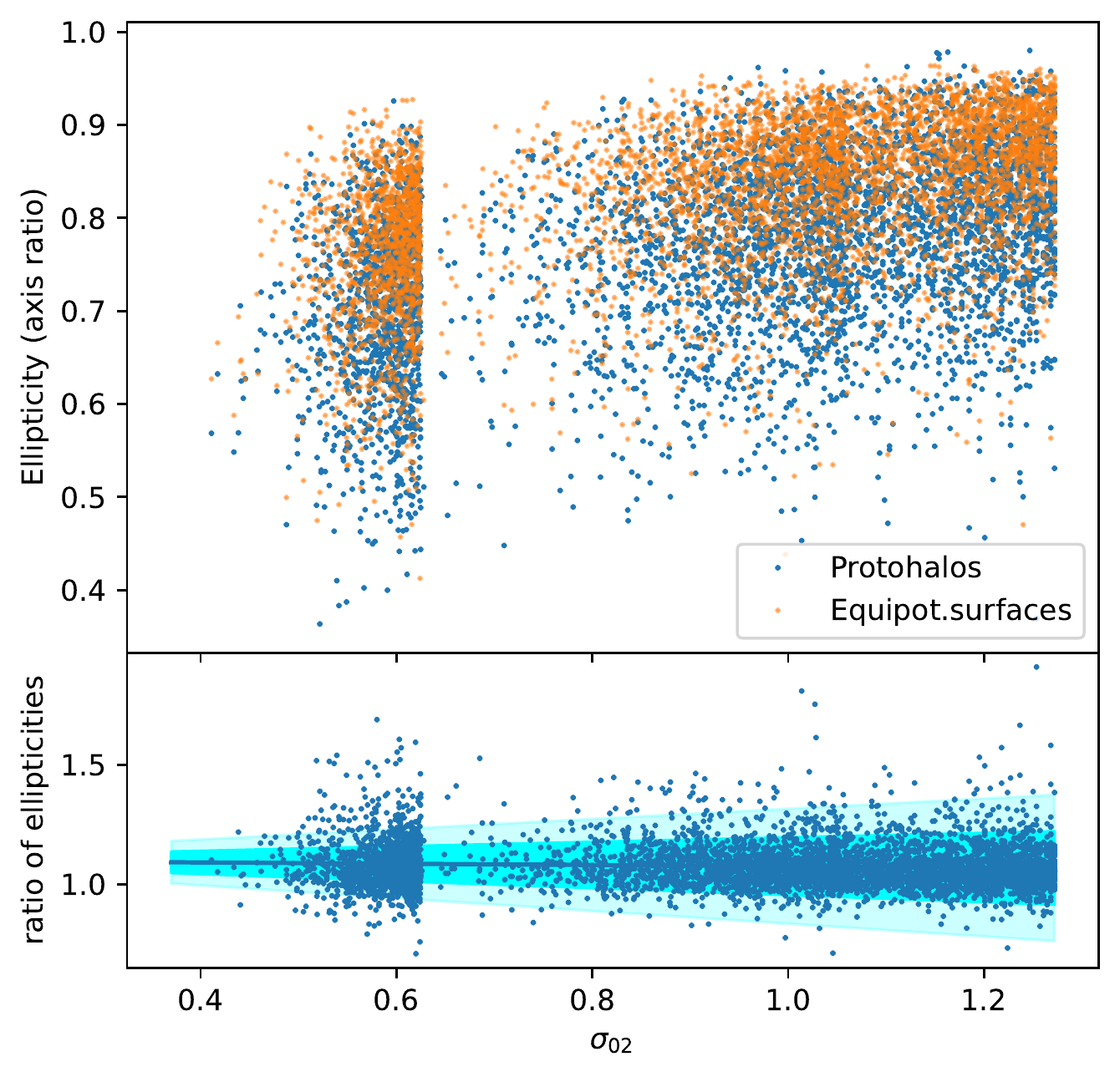}
    \caption{\emph{Top panel:} ellipticity of protohaloes (blue dots) vs. equipotential regions (orange). \emph{Bottom:} ratio of the two ellipticities. The  agreement is excellent, although equipotential surfaces are slightly more elliptical.}
    \label{fig:axisratios}
\end{figure}

\begin{figure}
    \centering
  \includegraphics[width=0.95\columnwidth]{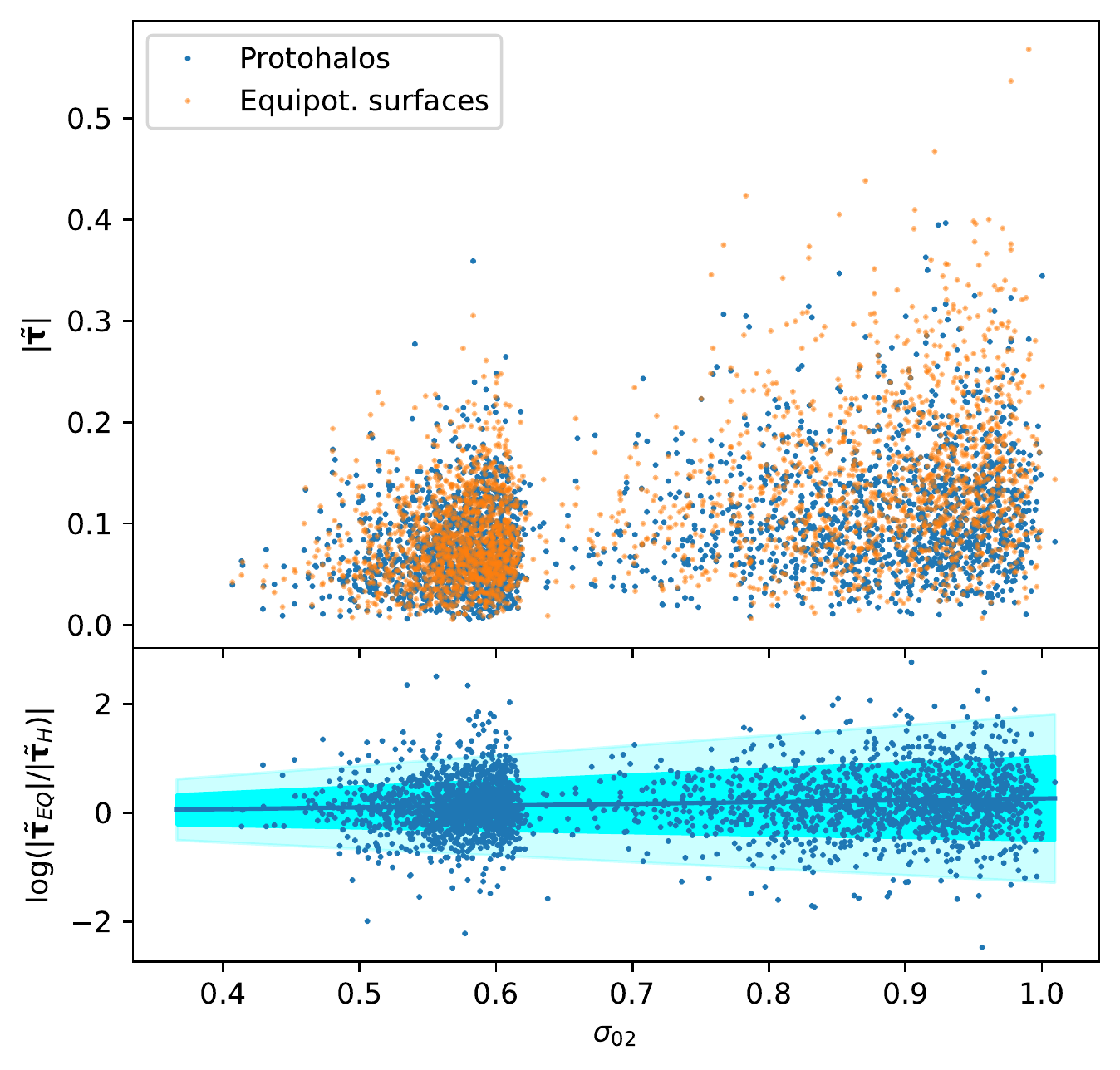}
    \caption{Reduced torque magnitudes (equation \eqref{eq:tau}) of actual protohaloes vs. equipotential surfaces as a function of $\sigma_{02}$ (largest masses on the left). \emph{Top panel:} torque of protohaloes (blue dots) and of equiptotential surfaces (orange dots). \emph{Bottom:} Logarithm of the ratio of the two magnitudes.}
    \label{fig:torques}
\end{figure}

For a more quantitative comparison, we 
quantify the overlap of the two regions in four different ways: \\
  (\emph{i}) the fraction of protohalo particles in the equipotential region (shown in the top panel of Figure~\ref{fig:cosines});\\
  (\emph{ii}) the matrix cosine between the inertia tensors $\mat{T}_{\mathrm{H}}$ and $\mat{T}_{\mathrm{EQ}}$ of each protohalo and equipotential region,
\begin{equation}
  \cos(\psi) \equiv
  \Tr(\mat{T}_{\mathrm{H}}\cdot \mat{T}_{\mathrm{EQ}})/\big[\Tr\big(\mat{T}_{\mathrm{H}}^2\big)\Tr\big(\mat{T}_{\mathrm{EQ}}^2\big)\big]^{1/2}\,
\label{eq:cosine}
\end{equation}
(shown in Figure~\ref{fig:cosines}, bottom), where
\begin{equation}
  T_{ij} = \sum_n [(\bm{r}-\rcm)_i(\bm{r}-\rcm)_j]_n\,;
\end{equation}
 (\emph{iii}) comparison of ellipticities $e_{\rm H}$ and $e_{\rm EQ}$, where $e\equiv \sqrt{1-a_3/a_1}$ and $a_3$ and $a_1$ are the smallest and largest eigenvalues of each inertia tensor (Figure~\ref{fig:axisratios}); and\\
 (\emph{iv}) comparison of the \emph{reduced} torques $\bm{\tilde \tau}_{\mathrm{H}}$ and $\bm{\tilde\tau}_{\mathrm{EQ}}$, defined as
\begin{equation}
    \tilde\tau_k \equiv \frac{\tau_k}{\tr(\mat{T})} = -\frac{1}{3}\, \varepsilon_{kij}\,\hat\epsilon_{ij}
    \label{eq:tau}
\end{equation}
with $\hat\epsilon_{ij}$ given in equation \eqref{eq:eij} (Figure~\ref{fig:torques}).  For \emph{(i-iii)}, the differences between the SO and EO halo sets were hardly noticeable, so we only show the (larger) SO sample.  However, the net torque on a sphere vanishes, so the estimated torque is quite sensitive to small changes in the halo (and hence protohalo) boundary.  So, for \emph{(iv)}, we show results for the EO sample.


Overall, protohalos tend to be slightly less elliptical than equipotential surfaces ($e_{\rm H} < e_{\rm EQ}$ in bottom panel of Figure~\ref{fig:axisratios}) but there is otherwise excellent agreement between the two. As a result, equipotential regions are more torqued (ratios in bottom panel of Figure~\ref{fig:torques} are slightly greater than unity; these ratios are slightly less than 1 for SO haloes). This small discrepancy is nearly mass independent.  

\begin{figure}
  \includegraphics[width=.9\columnwidth]{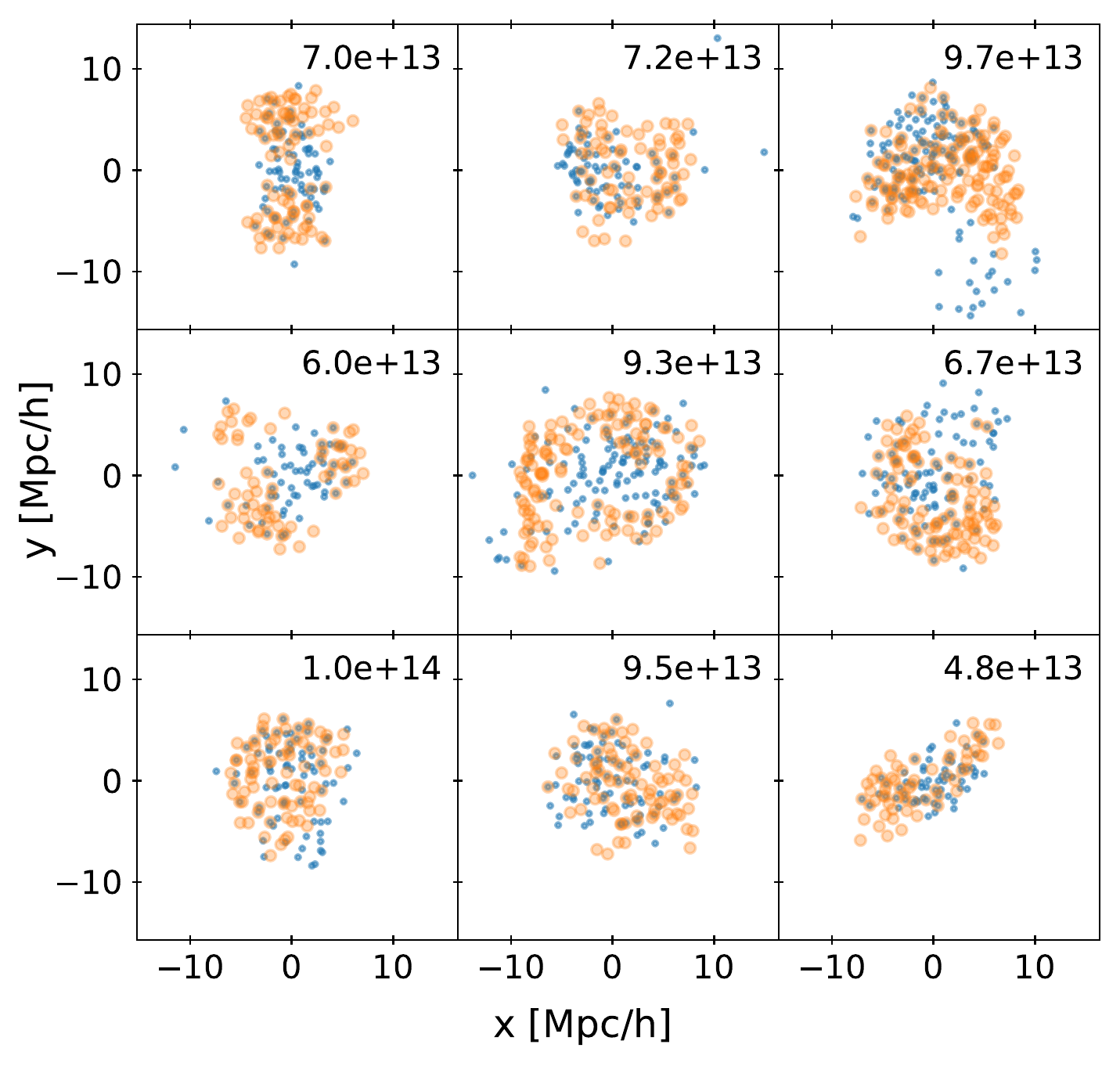}
\caption{\label{fig:fails} Same as Figure~\ref{fig:protohalos}, but for low mass objects that are outliers ($\leq 40\%$ of overlap) in the top panel of Figure~\ref{fig:cosines}.  Clearly, equipotential regions (orange symbols) do not identify the actual protohalo patches (blue).}
\end{figure}

To explore this further, Figure~\ref{fig:fails} shows a few of the objects with the smallest values (less than 0.4) in the top panel of Figure~\ref{fig:cosines}.  Whereas protohalo particles (blue) define a single blob, the equipotential regions are sometimes split into two or more components.  Clearly, our approach fails to describe these objects, suggesting that corrections beyond linear order in perturbations, which our approach ignores, can be important.  While it is certainly interesting to identify these corrections, they really only matter for a small fraction of the full halo sample, so we believe it is fair to conclude that, for the vast majority of halos, our approach identifies not just the protohalo center of mass but also its shape.

\section{Conclusions}\label{sec:conclude}

We have demonstrated that protohaloes are very well described as regions that maximize the energy overdensity $\epsilon$, bounded by surfaces of a suitably defined (equation~\ref{eq:f}) infall potential (Figure~\ref{fig:isosurfaces}). This approach captures both the centers of mass of protohaloes and their boundaries (Figures~\ref{fig:protohalos}--\ref{fig:torques}).

By connecting protohalo shapes to the infall pattern, our approach provides a framework not just for modeling halo shapes, but the anisotropy of infall velocity around clusters and the cross-correlation between dynamical and weak gravitational lensing mass estimates for clusters as well \citep[e.g.][]{wcs2010}, as well as how these correlate with the larger scale environment.

Going beyond a spherical model is crucial for predicting the initial torque.  Our minimum energy principle is therefore also suitable for making analytical predictions of the angular momentum of a protohalo patch.  This prediction uses the energy tensor $\epsilon_{ij}$ (equations~\ref{eq:eij} and~\ref{eq:tau}).  In future work, we will compare and contrast this with the predictions of Tidal Torque Theory \citep{doroshkevich1970, white1984, lp2001, ttt2002, cadiou2021}.

Although our approach proved to be robust to changes in how exactly halos are identified in the evolved field (e.g. SO vs EO halo finders), we believe that it can be most naturally compared with the objects found by the `boosted potential' halo finder recently proposed by \cite{boostpot}.  The two approaches are in fact very similar, with the important difference that ours (\emph{i}) derives equipotential surfaces from the energy minimisation principle, and (\emph{ii}) holds in the initial conditions, and therefore provides a natural framework for making analytical \emph{predictions} of halo statistics.  Moreover, as we noted, our equipotential approach encodes information about the assembly history of each object (Figure~\ref{fig:isosurfaces} and associated discussion); since it is similar in spirit to how halo substructures are identified in the boosted potential approach, a comparison of our predictions with such measurements should lead to interesting results, to be compared for instance with \cite{cadiou2020}.

\section*{Acknowledgements}
Thanks to Giulia Despali for sharing (and helping with) the Flora simulation, to Corentin Cadiou, Oliver Hahn, Dmitri Pogosyan for helpful comments, and to the organisers and other participants of the KITP Cosmic Web workshop. Our visit to KITP was supported by the National Science Foundation under PHY-1748958.

\section*{Data Availability}
The data of the Flora simulation, and the post-processing quantities used in this work, can be shared on reasonable request to the authors.



\bibliographystyle{mnras}
\bibliography{mybib}




\bsp	
\label{lastpage}
\end{document}